# Electronic stopping of slow protons in oxides: scaling properties

D. Roth[1], B. Bruckner[1], G. Undeutsch[1], V. Paneta[2], A.I. Mardare[3], C.L. McGahan[4], M. Dosmailov[5], J.I. Juaristi[6,7,8], M. Alducin[6,7], J.D. Pedarnig[5], R.F. Haglund, Jr.[4], D. Primetzhofer[2], and P. Bauer[1,6]

[1]Johannes-Kepler Universität Linz, IEP-AOP, Altenbergerstraße 69, A-4040 Linz, Austria

[2]Institutionen för Fysik och Astronomi, Uppsala Universitet, Box 516, S-751 20 Uppsala, Sweden

[3]Institut für Chemische Technologie Anorganischer Stoffe, Johannes-Kepler Universität Linz, Altenbergerstraße 69, A-4040 Linz, Austria

[4]Department of Physics and Astronomy, Vanderbilt University, Nashville, Tennessee 37235, USA

[5]Institut für Angewandte Physik, Johannes-Kepler Universität Linz, Altenbergerstraße 69, A-4040 Linz, Austria

[6]Donostia International Physics Center DIPC, P. Manuel de Lardizabal 4, 20018 Donostia-San Sebastián, Spain

[7]Centro de Física de Materiales CFM/MPC (CSIC-UPV/EHU), P. Manuel de Lardizabal 5, 20018 Donostia-San Sebastián, Spain

[8]Departamento de Física de Materiales, Facultad de Químicas, Universidad del País Vasco (UPV/EHU), Apartado 1072, 20018 Donostia-San Sebastián, Spain

**Abstract**

Electronic stopping of slow protons in ZnO, $VO_2$ (metal and semiconductor phases), $HfO_2$ and $Ta_2O_5$ was investigated experimentally. As a comparison of the resulting stopping cross sections (SCS) to data for $Al_2O_3$ and $SiO_2$ reveals, electronic stopping of slow protons does not correlate with electronic properties of the specific material such as band gap energies. Instead, the oxygen $2p$ states are decisive, as corroborated by DFT calculations of the electronic densities of states. Hence, at low ion velocities the SCS of an oxide primarily scales with its oxygen density.

Ions are slowed down in matter due to interaction with atomic nuclei and electrons; usually one differentiates between nuclear and electronic stopping. For many decades, fundamental research has been dedicated to accurate description of the relevant energy loss processes. The understanding gained is indispensable for wide-ranging applications – space research, material science, nuclear fusion and fission or radiation therapy [1]. In this context, a key quantity is the mean energy loss per path length, i.e., the stopping power $S = dE/dx$, with contributions due to electronic excitations, $S_e$, and nuclear collisions, $S_n$. To investigate the interaction of ions with compound materials, the stopping cross section per atom (SCS) $\varepsilon = S/n$ is a convenient measure, where $n$ denotes the atomic density of the target material.

At high ion velocities $v \gg v_F$ ($v_F$ denotes the Fermi velocity of the target electrons), $S_e$ is the main channel for energy loss of light ions in solids and accurate theoretical models are available [2, 3, 4]. At low ion velocities, $v \leq v_F$, $S_e$ is dominated by excitation of valence electrons, and also $S_n$ may contribute considerably to the overall energy dissipation rate. When the target electrons are described as a free electron gas (FEG) of effective density $n_e$, as characterized by the Wigner-Seitz radius $r_s = (3/4\pi n_e)^{1/3}$, electronic stopping due to electron-hole pair excitation by the screened ion charge is velocity-proportional, $S_e = Q(Z_1, r_s) \cdot v$ [5]. The friction coefficient $Q$ depends on $r_s$ and on the atomic number of the ion, $Z_1$. The nonlinear calculation of $Q$ [6] has been found to describe experimental proton stopping data quantitatively for metals and semiconductors for low velocities, up to $v \approx v_F$, using effective FEG electron densities $r_{s,\text{eff}}$, as derived from measured plasmon energies [7].

For very slow protons with $v \ll v_F$, deviations from velocity-proportionality of $S_e$ were reported for target materials featuring excitation thresholds in their electronic band structures; in noble metals, the $d$-bands exhibit an excitation threshold $E_d$ of several eV with respect to the Fermi energy $E_F$, so that at ion energies above ~ 1 keV, excitation of the $d$-bands becomes more and more effective and the stopping cross section rises with steeper slope [8, 9, 10, 11, 12]. Time-dependent density-functional theory (TD-DFT) calculations of $S_e$ of protons in Au confirmed this interpretation [13]. For large band gap insulators, electronic stopping was found to vanish below a threshold velocity $v_{th}$ – e.g., for LiF (band gap $E_{g,\text{LiF}} \approx 13.6$ eV) at velocities lower than $v_{th} \approx 0.1$ a.u. [14, 15]. Note that $v_{th}$ for LiF is considerably lower than the kink velocity for Au ($v_k \approx 0.2$ a.u.), even though $E_{g,\text{LiF}} \gg E_{d,\text{Au}} \approx 2$ eV. For LiF, TD-DFT calculations yielded considerably lower $S_e$ and a threshold velocity higher than the experimental value by a factor of ~ 2 [16]. Those calculations did, however, not allow for charge-exchange processes or defect production, which at grazing surface collisions had been identified as main channels of electronic losses [17, 18]. At smaller impact parameters, electron promotion in atomic collisions was suggested to be responsible for the efficient electronic stopping of protons in ionic insulators [19]. Nevertheless, a conclusive description is still missing.

The interplay between band gaps and electronic stopping of slow ions is complex, since $E_g$ will be modified by the electric field of the ion [20, 21]. Recently, the importance of static crystal effects (momentum transfer from the crystal) [22] and dynamic defect states ("electron elevator") [23] has been revealed. In the band gap of Si dynamic defect states induced by a *moving* Si ion lead to a very efficient transfer of electrons from the valence band to the conduction band. This mechanism is expected to be relevant also for stopping of slow protons [23].

Metal oxides exist in building blocks of different sizes with widely differing numbers of valence electrons per building block, $N_{val}$, and exhibit more covalent bonds with smaller band gaps as compared to alkali halides [24]. In this study, we investigate systematically how in oxides proton stopping is influenced by electronic features, such as $E_g$ or the valence electron density. To this aim, we studied electronic stopping of protons in $ZnO$, $VO_2$, $HfO_2$ and $Ta_2O_5$ in the range $0.15$ a.u. $\leq v \leq 0.64$ a.u. ($500$ eV $-$ $10$ keV protons) and relate these results to those obtained for $SiO_2$ [14] and $Al_2O_3$ [21]. In this context, $VO_2$ is a key material to investigate the influence of $E_g$ on $S_e$, due to the insulator-to-metal transition at a critical temperature $T_C \approx 67$ °C [25, 26]. The selection of oxides was made to cover wide ranges of band gaps, $0$ eV $\leq E_g \leq 9$ eV, and one to five oxygen atoms per building block, corresponding to $6 \leq N_{val} \leq 30$, equivalent to valence electron densities corresponding to $1.57 \leq r_s \leq 1.86$ (see Tab. 1).

| Oxide | $E_g$ [eV] | $N_{val}$ | $r_s$ [a.u.] | $\varepsilon_{ox,expt}$ (0.2 a.u.) [$10^{-15}$eVcm$^2$/atom] |
|---|---|---|---|---|
| VO$_2$ | 0 … 0.7 | 13 | 1.58 | 2.33 |
| ZnO | 3.4 | 6 | 1.86 | 1.60 |
| Ta$_2$O$_5$ | 3.9 | 30 | 1.69 | 2.71 |
| HfO$_2$ | 5.5 | 12 | 1.69 | 2.62 |
| Al$_2$O$_3$ | 8 | 18 | 1.57 | 2.00 |
| SiO$_2$ | 9 | 12 | 1.72 | 2.10 |

TAB. 1 *Electronic properties of ZnO, VO$_2$, HfO$_2$ SiO$_2$ Al$_2$O$_3$ and Ta$_2$O$_5$: $E_g$, $N_{val}$, $r_s$ derived from $n_{val} = N_{val} \cdot n$, and experimental values $\varepsilon_{ox,expt}$ for v = 0.2 a.u. (1 keV protons).*

The experiments were performed at the IEP in Linz employing the UHV time-of-flight low energy ion scattering (TOF-LEIS) setup ACOLISSA [27]. All samples were prepared *ex-situ*; HfO$_2$ thin films were deposited on SiO$_2$/Si by atomic layer deposition [28]; VO$_2$ thin films were sputter deposited on Si and subsequently thermally oxidized [29]. The annealed VO$_2$ films were checked for the first-order phase transition by optical transmission (near infrared) while cycling forward and backward through $T_C$. Ta$_2$O$_5$ samples were produced by anodization of a Ta sheet [30, 31], and ZnO samples were prepared in three different ways: thermal oxidation in air of a high purity Zn sheet, pulsed laser deposition on PET [32], and sputter deposition on glass. Time-of-flight elastic recoil detection (TOF-ERD) measurements at Uppsala University yielded the expected stoichiometry and impurity concentrations below ~ 2 %. Time-of-flight medium energy ion scattering (TOF-MEIS) [33] was employed to check the homogeneity of thin film samples. TOF-LEIS spectra were recorded using hydrogen and deuterium beams (monomers and dimers) in the range of 0.5 keV/u – 10 keV/u. The projectiles impinge at normal incidence and probe the bulk properties in a depth of at least several nanometers; at a scattering angle $\theta = 129$ °, time of flight (TOF) is measured for backscattered projectiles of any charge state. From the energy-converted spectra, the electronic SCS per atom, $\varepsilon_{ox}$, was deduced.

For nanometer films, thickness was determined by Rutherford backscattering spectrometry (RBS). To evaluate $\varepsilon_{ox}$, experimental spectrum widths were compared to corresponding Monte Carlo simulations (TRBS, [34]), in order to disentangle electronic and nuclear stopping. In the simulations, a screened Coulomb potential (ZBL, [35]) was used to handle scattering in close and distant collisions; $\varepsilon_{ox}$ was optimized to reproduce the width of the experimental spectrum ([11]).

For ZnO, $Ta_2O_5$, and at very low ion velocities $\varepsilon_{ox}$ was deduced from the height ratio of energy spectra, $H_{ox}/H_{ref}$, recorded for the oxide and a reference sample (of known SCS $\varepsilon_{ref}$) for the same primary charge (similarly as in [36]). The reference targets (polycrystalline Cu and Au) were cleaned employing 3 keV $Ar^+$ sputtering; surface purity was checked by Auger electron spectroscopy (AES). The experimental height ratios were compared to results from corresponding TRBS simulations; $\varepsilon_{ox}$ was evaluated close to the high energy edge of the spectrum, where the shapes of the experimental spectra are perfectly reproduced [37]. In the simulations, $\varepsilon_{ox}$ is the only parameter to be optimized.

The statistical uncertainties of $\varepsilon_{ox}$ range from < 7 % (evaluation of spectrum widths) to 10 % - 15 % (evaluation of spectrum heights), with highest uncertainties at lowest ion velocities. Systematic errors due to $\varepsilon_{ref}$ or thin film thickness determination, and the interaction potential in the simulations are < 10 %.

In Fig. 1, $\varepsilon_{ox}$ is shown for metallic and semiconducting $VO_2$ for H ions (protons and deuterons), measured at 300 K and 373 K, respectively, with identical $\varepsilon_{ox}$-values for both phases. Thus, in $VO_2$ proton stopping [38] is independent of the existence of a band gap, $E_g$. This finding is corroborated when comparing the results for $VO_2$, $HfO_2$ ($E_g \approx 5.5$ eV) and $SiO_2$ ($E_g \approx 9$ eV), since their $\varepsilon_{ox}$ data coincide within experimental uncertainty (see Fig. 2a). Apparently, for the investigated binary oxides, band gaps are irrelevant for electronic stopping even at low ion velocities, where energy transfers in a ion-electron collisions are small.

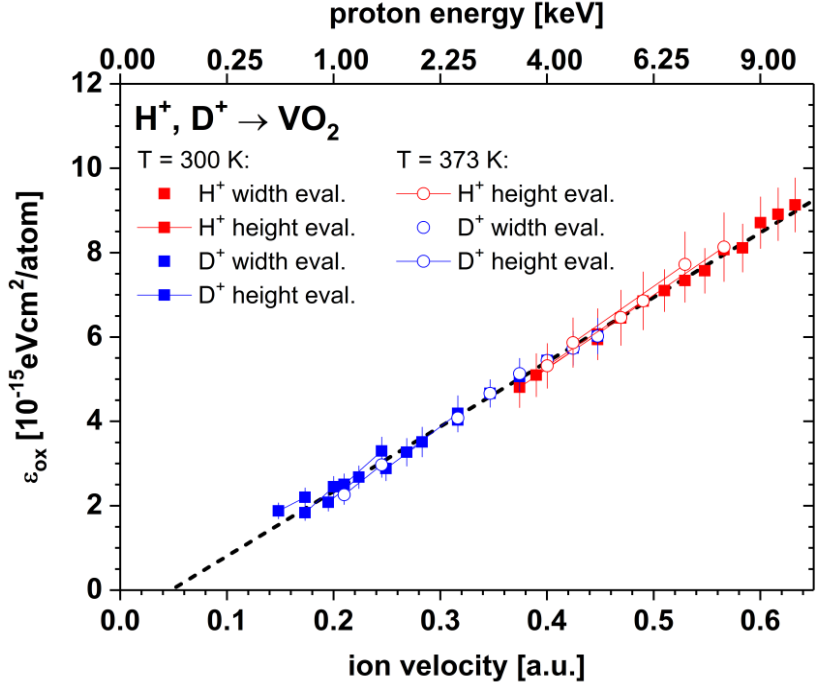

FIG. 1 $\varepsilon_{ox}$ *of VO$_2$ for H ions (protons and deuterons) in both, metallic (open symbols) and semiconducting phases (full symbols) are shown as function of the ion velocity. Evaluation of both, widths and heights of the spectra, yields concordant results (see legend). The upper labelling of the abscissa denotes the corresponding proton energy.*

For oxides, there is no simple correspondence between valence electron densities, plasmon energies and electronic stopping: e.g., for ZnO the experimental plasmon energy [39] is consistent with the valence electron density, while for protons $\varepsilon_{ox}$ is lower by a factor of ~ 2 than anticipated for a FEG [6] (for $v = 0.2$ a.u.). Moreover, the SCS data spread much more than one would anticipate from their $r_s$ values (see Tab. 1, [6]). Clearly, it does not make sense to describe oxides as FEG.

While it is easily possible at high ion velocities to relate electronic stopping of a compound, $\varepsilon_{A_xB_{1-x}}$, to the SCS of the constituents, $\varepsilon_A$ and $\varepsilon_B$, by applying Bragg's rule [40], $\varepsilon_{A_xB_{1-x}} = x \cdot \varepsilon_A +(1-x) \cdot \varepsilon_B$, this is a doubtful approach at low ion velocities, where formation of a compound changes the valence electron states considerably. The breakdown of the additivity rule can be seen in Fig. 2a, where the low velocity SCS of selected oxides (ZnO, VO$_2$, SiO$_2$, HfO$_2$, Al$_2$O$_3$, Ta$_2$O$_5$) are presented together with Bragg's rule predictions using data from [36, 41, 42, 43]: the additivity rule results are higher by more than a factor of 2 at lowest velocities, with largest discrepancies for ZnO and SiO$_2$.

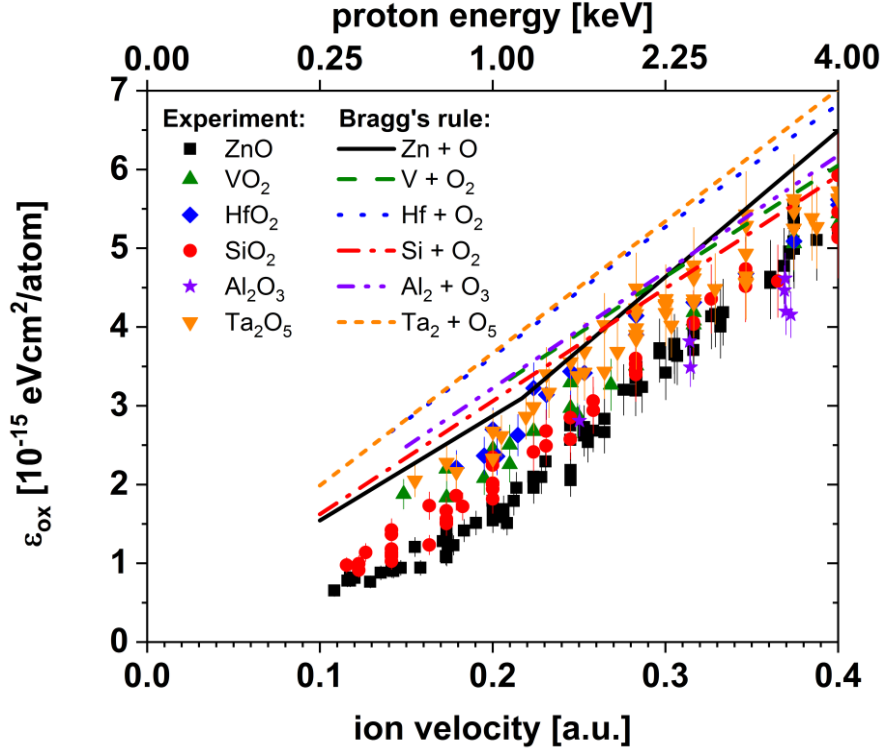

FIG. 2a) *For ZnO, VO$_2$, Ta$_2$O$_5$, HfO$_2$, Al$_2$O$_3$ [21], and SiO$_2$ [14], $\varepsilon_{ox}$ are displayed as a function of the ion velocity (full symbols), together with Bragg's rule predictions using data from [36, 41, 42, 43].*

In Fig. 2b, we present our results for the oxides as SCS per oxygen atom, $\varepsilon_O = \varepsilon_{ox} \cdot (1-x)$, i.e., we relate $\varepsilon_{ox}$ to the oxygen sub-lattice. In this way, all data coincide within experimental uncertainties, except for ZnO, for which $\varepsilon_O$ rises with steeper slope at $v \geq 0.25$ a.u., due to the contribution from the full *d*-band, similarly as for metallic Zn [41]. In fact, in oxides A$_x$O$_{1-x}$, the SCS is proportional to the atomic fraction of oxygen, 1-*x*, while detailed electronic properties such as band gap energy or valence electron density are not relevant. At higher ion velocities, such a behavior has been observed for Al$_2$O$_3$, SiO$_2$ and H$_2$O ice [44] as well as for HfO$_2$ versus SiO$_2$ [28] and traced back to an O $2p^6$ configuration as if in oxides the ionic character of the local bonds would prevail. At low ion velocities, however, details of the density of states (DOS) might be highly relevant, since even the subtle differences between specific metals have clear impact on the observed $S_e$, e.g., for Au and Pt [10, 11].

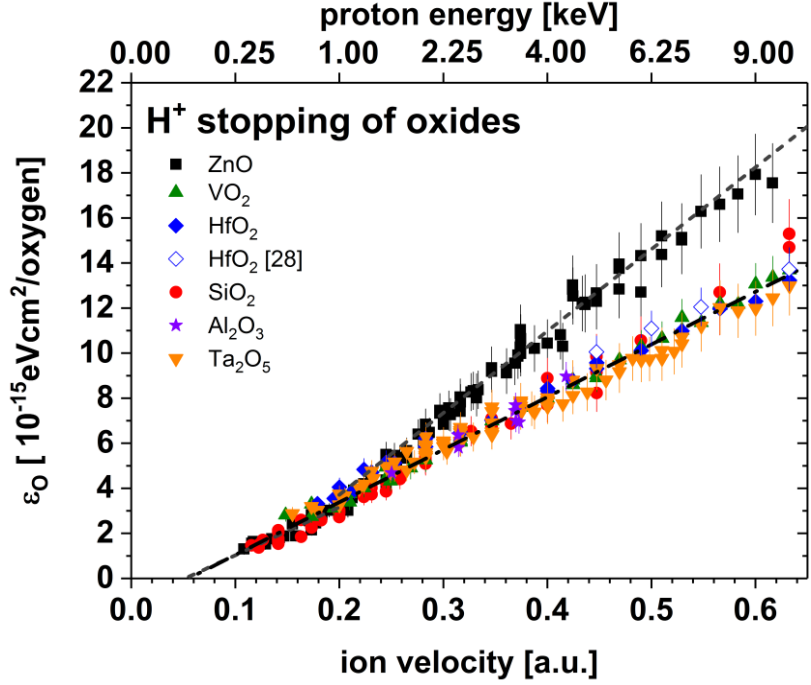

Fig. 2b) *The experimental data of Fig. 2a are shown as SCS per O atom, $\varepsilon_O = \varepsilon_{ox}\cdot(1-x)$, in a wider velocity range. For $HfO_2$, also data from Ref. [28] are shown which exhibit excellent agreement with the present results. The upper labelling of the abscissa denotes the corresponding proton energy.*

In order to obtain quantitative information on the unperturbed electronic density of states (DOS) of all presented oxides, DFT calculations were performed with the VASP code [45, 46]. For the metallic rutile structure of $VO_2$ the PBE exchange correlation functional was used [47]. For the monoclinic structure of $VO_2$ we used a PBE + U approach with U = 3.5 eV, in accordance with, e.g., [48]. For all other oxides, the hybrid PBE0 exchange correlation functional was employed [49]. In all calculations, the energy cutoff for the plane wave basis sets is 400 eV and projector augmented wave (PAW) potentials were utilized [50, 51]. The Brillouin zone is sampled by a $11 \times 11 \times 11$ Monkhorst-Pack grid of **k**-points [52] for monoclinic $VO_2$, a $7 \times 7 \times 11$ grid for rutile $VO_2$, and a $7 \times 7 \times 7$ grid for the other oxides. In evaluating the DOS, the occupancies of the electronic states are determined with the tetrahedron method.

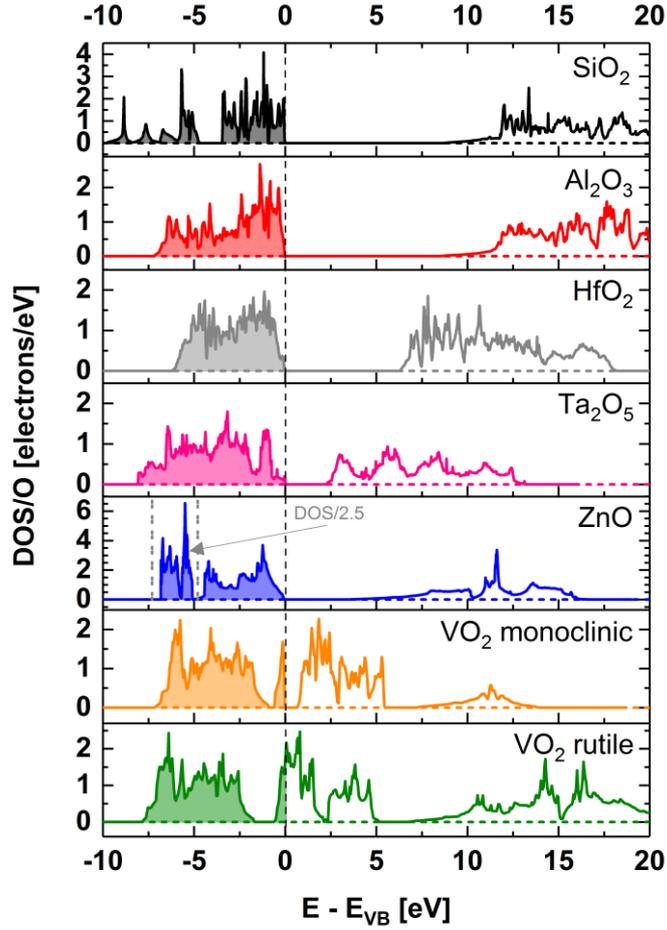

FIG. 3 *DOS per O atom for selected oxides (for VO$_2$ both metallic rutile and semiconducting monoclinic phases) are depicted as function of E - E$_{VB}$. For ZnO, the high DOS below − 4.8 eV (due to d electrons of Zn) has been scaled down by a factor 2.5. "Zero" unoccupied DOS at higher energies is due to a limited number of bands used in the calculations.*

The results for the oxides of interest in terms of DOS per oxygen atom (DOS/O) are shown as function of $E - E_{VB}$ in Fig. 3, where $E_{VB}$ represents the highest occupied state of the valence band. Integration of the DOS/O from − 10 eV up to $E_{VB}$ yields ~ 6 electrons for all oxides, with the exception of ZnO (contribution from *d* electrons of Zn). Integration of the unoccupied DOS/O in an interval of 10 eV beyond the band gap is rather independent of the metal/semiconductor atom (4 to 6 electrons). Thus, the observed scaling properties of $\varepsilon_O$ may be interpreted in a similar way as electronic stopping of protons in metals [36]. Another aspect of these results is that for stopping of slow protons in an oxide A$_x$O$_{1-x}$, Bragg's rule is simplified since the contribution of the cations, $\varepsilon_A$, can be set to zero.

In contrast to metals, however, a linear fit to $\varepsilon_O$ yields an apparent velocity threshold of $v_{th} \approx 0.055$ a.u. (even for the metallic phase of VO$_2$), independent of the (unperturbed) band gap. It is not yet clear how the existence of such an apparent velocity threshold should be interpreted [20, 21, 22, 23]. In fact, $v_{th} \approx 0.055$ a.u. is comparable to the values observed for ionic crystals such as LiF ($v_{th} \approx 0.1$ a.u.) or KCl ($v_{th} \approx 0.07$ a.u.) [14], and for a covalent semiconductor like Ge ($v_{th} \approx 0.026$ a.u.) [37, 53]. It remains unclear, whether the lack of correlation between $v_{th}$ and $E_g$ points towards Coulomb collisions with electrons in a strongly perturbed band, towards a different process like the "electron elevator", or towards electron promotion in an atomic collision. In any case, the energy loss mechanism appears to be similar for all oxides. It may be interesting to compare the response of the electronic system to energy deposition by a slow ion and by laser pulses: when exposed to high power femtosecond laser pulses the band gap in VO$_2$ collapses instantaneously [54].

To conclude, we present electronic stopping data $\varepsilon_{ox}$ for slow protons in selected oxides with a wide range of electronic properties, e.g., band gaps from 0 eV up to 9 eV, and 6 to 30 valence electrons per building block. Our data reveal that $\varepsilon_{ox}$ is independent of $E_g$, but scales with the atomic fraction of oxygen in the building block, since all oxides studied exhibit ~ 6 valence electrons per O atom, as corroborated by DFT-calculations of the electronic DOS – even if the chemical bonds are only partly ionic. The irrelevance of $E_g$ may be either due to a strong modification of the electronic band structure or to dynamic defect states induced in the band gap by the ion – causing a locally reduced band gap ("metallization"). Nevertheless, to describe the valence electrons in the oxides as a FEG of effective density is not an expedient approach. In any case, the present results permit to fix the electronic stopping of any oxide of interest. This is important, for instance, when estimating the electron yield emitted from the first wall of a nuclear fusion device, or to determine the mean range of slow protons in an oxide. Another observation is that our $\varepsilon_O$ data extrapolate to an apparent velocity threshold, $v_{th} \approx 0.055$ a.u., even for the metallic phase of VO$_2$ - it simply seems to be an oxygen property. Definite answers require theoretical models with realistic description of ion-electron interactions inside band gap materials.

Financial support of this work by the FWF (FWF-Project No. P22587-N20 and FWF-Project No. P25704-N20) is gratefully acknowledged. MA and JIJ acknowledge financial support by the Gobierno Vasco-UPV/EHU project IT756-13, and the Spanish Ministerio de Economa y Competitividad (Grants No. FIS2013-48286-C02-02-P and FIS2016-76471-P). Fabrication and characterization of VO$_2$ films at Vanderbilt University (CMG and RFH) was supported by a grant from the National Science Foundation (DMR-1207507). A research infrastructure fellowship of the Swedish Foundation for Strategic Research (SSF) under Contract No. RIF14-0053 supporting accelerator operation is acknowledged. PB expresses his gratitude for the kind hospitality at the DIPC in San Sebastián. We are grateful to Len Feldman, Pedro Echenique, Andres Arnau and Peter Zeppenfeld for inspiring discussions.

**References and remarks:**

[1] A. Vantomme, Nucl. Instr. Meth. B **371**, 12-26 (2016).

[2] C.P. Race, D.R. Mason, M.W. Finnis, W.M.C. Foulkes, A.P. Horsfield, and A.P. Sutton, Rep. Prog. Phys **73**, 116501 (2010).

[3] P. Sigmund, *Particle Penetration and Radiation Effects – General Aspects and Stopping of Swift Point Charges* (Springer Verlag, Berlin-Heidelberg, 2006).

[4] J.F. Ziegler, J. Appl. Phys. **85**, 1249 (1999).

[5] E. Fermi and E. Teller, Phys. Rev **72**, 399-408 (1947).

[6] P.M. Echenique, R.M. Nieminen, and R.H. Ritchie, Solid State Commun. **37** (10), 779-781 (1981).

[7] A. Mann and W. Brandt, Phys. Rev. B **24**, 4999-5003 (1981).

[8] Velocities and lengths are given in atomic units: $v_0$ = c/137 (c denotes the speed of light), and the Bohr radius $a_0$ = 0.529Å, respectively. For protons with a kinetic energy of 25 keV, $v = v_0$.

[9] J.E. Valdes, J.C: Eckardt, G.H. Lantschner, and N.R. Arista, Phys. Rev. A **49**, 1083-1088 (1994).

[10] S.N. Markin, D. Primetzhofer, M. Spitz, and P. Bauer, Phys. Rev. B **80**, 205105 (2009).

[11] D. Goebl, D. Roth, and P. Bauer, Phys. Rev. A **87**, 062903 (2013).

[12] E.D. Cantero, G.H. Lantschner, J.C. Eckardt, and N.R. Arista, Phys. Rev. A **80**, 032904 (2009).

[13] M.A. Zeb, J. Kohanoff, D. Sánchez-Portal, A. Arnau, J.I. Juaristi, and E. Artacho, Phys. Rev. Lett. **108**, 225504 (2012).

[14] S.N. Markin, D. Primetzhofer, and P. Bauer, Phys. Rev. Lett. **103**, 113201 (2009).

[15] L.N. Serkovic Loli, E.A. Sánchez, O. Grizzi, and N.R. Arista, Phys. Rev. A **81**, 022902 (2010).

[16] J.M. Pruneda, D. Sánchez-Portal, A. Arnau, J.I. Juaristi, and E. Artacho, Phys. Rev. Lett. **99**, 235501 (2007).

[17] C. Auth, A. Mertens, H. Winter, and A. Borisov, Phys. Rev. Lett. **81,** 4831 (1998).

[18] P. Roncin, J. Villette, J. P. Atanas, and H. Khemliche, Phys. Rev. Lett. **83**, 864 (1999).

[19] P.A. Zeijlmans van Emmichoven, A. Niehaus, P. Stracke, F. Wiegershaus, S. Krischok, V. Kempter, A. Arnau, F.J. Garcia de Abajo, and M. Penalba, Phys. Rev. B **59**, 10950 (1999).

[20] B. Solleder, L. Wirtz, and J. Burgdörfer, Phys. Rev. B **79**, 125107 (2009).

[21] K. Eder, D. Semrad, P. Bauer, R. Golser, P. Maier-Komor, F. Aumayr, M. Peñalba, A. Arnau, J.M. Ugalde, and P.M. Echenique, Phys. Rev. Lett. **79**, 4112 (1997).

[22] E. Artacho, J. Phys. Condens. Matter **19**, 275211 (2007).

[23] A. Lim, W.M.C. Foulkes, A.P. Horsfield, D.R. Mason, A. Schleife, E.W. Draeger, and A.A. Correa, Phys. Rev. Lett. **116**, 043201 (2016).

[24] W.D. Grobman, D.E. Eastman, and M.L. Cohen, Phys. Lett. **43A**, 1, 49-50 (1973).

[25] F.J. Morin, Phys. Rev. Lett. **3**, 34 (1959).

[26] C. Weber, D.D. O'Regan, N.D.M. Hine, M.C. Payne, G. Kotliar, and P.B. Littlewood, Phys. Rev. Lett. **108**, 256402 (2012).

[27] M. Draxler, S.N. Markin, S.N. Ermolov, K. Schmid, C. Hesch, R. Gruber, A. Poschacher, M. Bergsmann, and P. Bauer, Vacuum **73**, 39-45 (2004).

[28] D. Primetzhofer, Phys. Rev. A **89**, 032711 (2014).

[29] R.E. Marvel, R.R. Harl, V. Craciun, B.R. Rogers, and R.F. Haglund, Jr., Acta Materialia **91**, 217-226 (2015).

[30] A.W. Hassel and D. Diesing, Thin Solid Films **414**, 296-303 (2002).

[31] J.P. Kollender, M. Voith, S. Schneiderbauer, A.I. Mardare, and A.W. Hassel, J. Electroanal. Chem. **740**, 53-60 (2015).

[32] M. Dosmailov, L.N. Leonat, J. Patek, D. Roth, P. Bauer, M.C. Scharber, N.S. Sariciftci, and J.D. Pedarnig, Thin Solid Films **591**, 97-104 (2015).

[33] M.K. Linnarsson, A. Hallén, J. Åström, D. Primetzhofer, S. Legendre, and G. Possnert, Rev. Sci. Instr. **83**, 095107 (2012).

[34] J.P. Biersack, E. Steinbauer, and P. Bauer, Nucl. Instr. Meth. B **61**, 77-82 (1991).

[35] J.F. Ziegler, J.P. Biersack and U. Littmark, *The stopping and range of ions in solids*, (Pergamon Press, New York, 1985), Vol. 1.

[36] D. Roth, B. Bruckner, M.V. Moro, S. Gruber, D. Goebl, J.I. Juaristi, M. Alducin, R. Steinberger, J. Duchoslav, D. Primetzhofer, and P. Bauer, Phys. Rev. Lett. **118**, 103401 (2017).

[37] D. Roth, D. Goebl, D. Primetzhofer, and P. Bauer, Nucl. Instr. Meth. B **317**, 61-65 (2013).

[38] Since no isotope or vicinage effects are observed, we refer in the discussion for the projectiles as to protons.

[39] R.L. Hengehold, R.J. Almassy, and F.L. Predrotti, Phys. Rev. B **1**, 4784-4791 (1970).

[40] W.H. Bragg and R. Kleeman, Phil. Mag. **10**, 318 (1905).

[41] D. Goebl, W. Roessler, D. Roth, and P. Bauer, Phys. Rev. A **90**, 042706 (2014).

[42] D. Primetzhofer, S. Rund, D. Roth, D. Goebl, and P. Bauer, Phys. Rev. Lett. **107**, 163201 (2011).

[43] J. F. Ziegler, J. P. Biersack, and M. D. Ziegler, *SRIM, The Stopping and Range of Ions in Matter*, (SRIM Company, Chester, 2008).

[44] P. Bauer, R. Golser, F. Aumayr, D. Semrad, A. Arnau, E. Zarate, and R. Diez-Muiño, Nucl. Instr. Meth. B **125**, 102-105 (1997).

[45] G. Kresse and J. Furthmüller, Comput. Mater. Sci. **6**, 15 (1996).

[46] G. Kresse and J. Furthmüller, Phys. Rev. B. **54**, 11169 (1996).

[47] J.P. Perdew, K. Burke, and M. Ernzerhof, Phys. Rev. Lett. **77**, 3865 (1996).

[48] B.Y. Qu, H.Y. He, and B.C. Pan, J. Appl. Phys. **110**, 113517 (2011).

[49] C. Adamo and V. Barone, J. Chem. Phys. **110**, 6158 (1999).

[50] P.E. Blöchl, Phys. Rev. B **50**, 17953 (1994).

[51] G. Kresse and D. Joubert, Phys. Rev. B **59**, 1758 (1999).

[52] H.J. Monkhorst and J.D. Pack, Phys. Rev. B **13**, 5188 (1976).

[53] R. Ullah, F. Corsetti, D. Sánchez-Portal, and E. Artacho, Phys. Rev. B **91**, 125203 (2015).

[54] D. Wegkamp, M. Herzog, L. Xian, M. Gatti, P. Cudazzo, C.L. McGahan, R.E. Marvel, R.F. Haglund, Jr., A. Rubio, M. Wolf, and J. Stähler, Phys. Rev. Lett. **113**, 216401 (2014).